# OBJECTS IN TELESCOPE ARE FARTHER THAN THEY APPEAR

## How diffraction tricked Galileo into mismeasuring the distances to the stars.

Christopher M. Graney
Jefferson Community College
1000 Community College Drive
Louisville, KY 40272
christopher.graney@kctcs.edu
www.jefferson.kctcs.edu/faculty/graney
502-213-7292

ABSTRACT

Galileo determined distances to stars based on the assumption that stars were suns, the apparent sizes of stars as seen through his telescope, and basic geometry. However, the apparent sizes that he measured were the result of diffraction and not related to the actual sizes of the stars. Galileo's methods and observations were good, but since he was unknowingly observing diffraction artifacts and not the physical bodies of stars he greatly underestimated the distances to the stars.

The wave nature of light is not part of students' common experiences, so often physics teachers and text books will add an historical anecdote about how scientists, too, were tricked by light. A common one is how in the early 19th century Poisson declared that since Fresnel's ideas on the wave nature of light implied that the shadow cast by a disk would contain a bright spot at it's center, Fresnel's ideas were obviously flawed. The spot was later detected, proving Fresnel right! But recent studies of Galileo's work have brought to light a story about diffraction that may displace Poisson's spot as favored historical anecdote, for it seems that diffraction tricked Galileo, too. Diffraction of light caused Galileo to mismeasure the distances to the stars.

Throughout his career Galileo held the view that the stars were suns located at vast distances from Earth -- a view that he discusses in depth on the "Third Day" of his *Dialogue Concerning the Two Chief World Systems*. For example, in arguing for the motion of Earth and the lack of motion of the Sun, he states, "See then, how neatly the precipitous motion of each twenty-four hours is taken away from the universe, and how the fixed stars (which are so many suns) agree with our sun in enjoying perpetual rest."[1] Galileo argued that with a good telescope one could measure the angular sizes of stars, and that the stars typically measured a few arc-seconds[2] in diameter.[3] He felt it was possible in the case of bright stars to independently confirm the sizes measured through a telescope via clever naked-eye measurements.[4] And if the stars are suns, and if it is possible to measure their sizes, then it is possible to use basic geometry to determine their distances:

> ...the apparent diameter of the sun.. is about one-half a degree, or 30 minutes; this is 1800 seconds, or 108,000 third-order divisions. Since the apparent diameter of a fixed star of the first magnitude [a bright star to the naked eye] is no more than 5 seconds, or 300 thirds, and the diameter of one of the sixth magnitude [a very dim star as seen with the naked eye] measures 50 thirds... then the diameter of the sun contains the diameter of a fixed star of the sixth magnitude 2,160 times. Therefore if one assumes that a fixed star of the sixth magnitude is really equal to the sun and not larger.... the distance of a fixed star of sixth magnitude is 2,160 radii of the Earth's orbit.[5]

In modern terms we would call this distance 2,160 astronomical units (AU). By this same method, bright stars with apparent diameters of 5 arc-seconds lie at approximately 360 AU. So according to Galileo the stars we can see range from being hundreds to thousands of AU distant. This is pretty far -- Neptune lies approximately 30 AU from the Sun -- but today we know that stars are vastly more distant than Galileo figured. The nearest stars are almost 300,000 AU distant.

It might seem like Galileo was simply making assumptions about distances to drive home a point, but recent work seems to indicate that Galileo really could measure tiny sizes with his telescopes. We now know that Galileo developed an ingenious technique for making measurements with a telescope that allowed him to measure Jupiter's apparent diameter as being 41.5 arc-seconds one month, and then to notice that the diameter decreased (as the distance between Earth and Jupiter increased) to 39.25 arc-seconds a few months later.[6] We know he could accurately plot the position of an object as faint as Neptune.[7] It now seems clear he could generally measure positions and sizes of small objects down to arc-second accuracy[8], and he was making these kinds of measurements and doing these kinds of calculations over a span of many years[9]. He could certainly back up his assumptions about star sizes with data.

So where would that data have come from? Why would Galileo think that bright stars have apparent diameters of about 5 arc-seconds and that size drops with magnitude (brightness) down to dim stars having diameters of about 5/6 arc-seconds? Because, thanks to diffraction, that's what he saw through his telescope.

Because stars are so far away and thus so apparently tiny, the light from a star passing through the lens of a telescope is a textbook case of light from a point source diffracting through a circular aperture. The magnified image of a star seen in a telescope is not the star's physical body but rather a diffraction pattern. Plenty of physics texts include the diffraction pattern for a circular aperture (Figure 1) and the equation for the radius of the central maximum in the pattern, known as the Airy Disk radius ($\theta_A$)

$$\theta_A = 1.22\lambda/D.$$

Here $\lambda$ is the wavelength of light and D is the diameter of

FIGURE 1

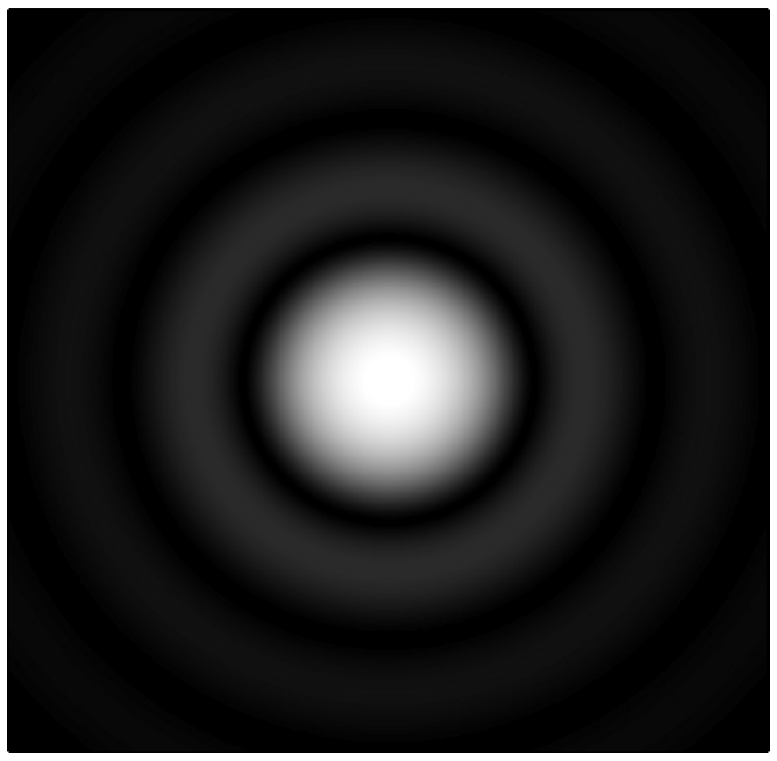

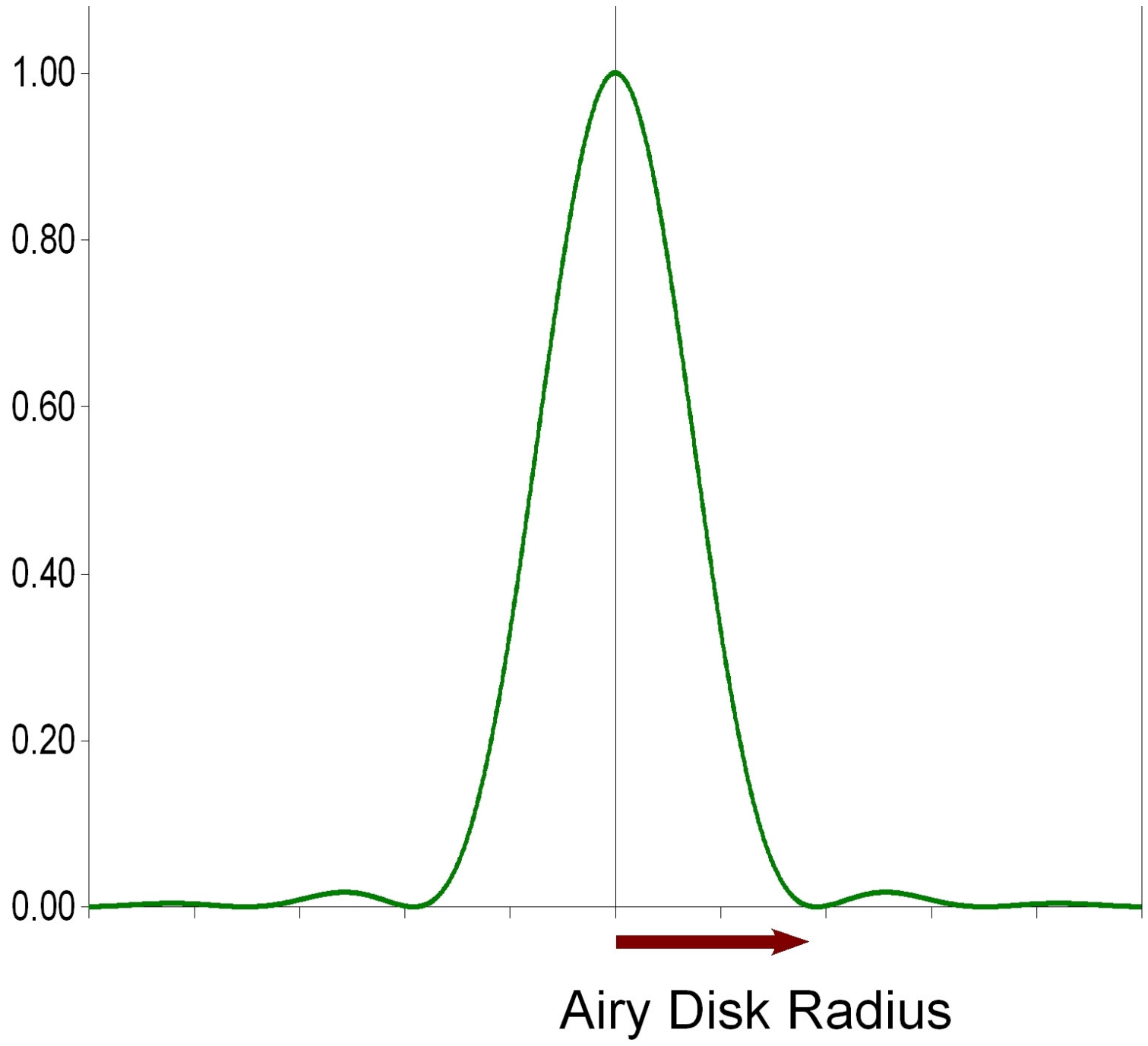

**Above --**
**Circular aperture diffraction pattern.**

**Top right --**
**Intensity plot for diffraction pattern, showing Airy Disk radius.**

**Bottom right --**
**Intensity plot for diffraction pattern (log axis).**

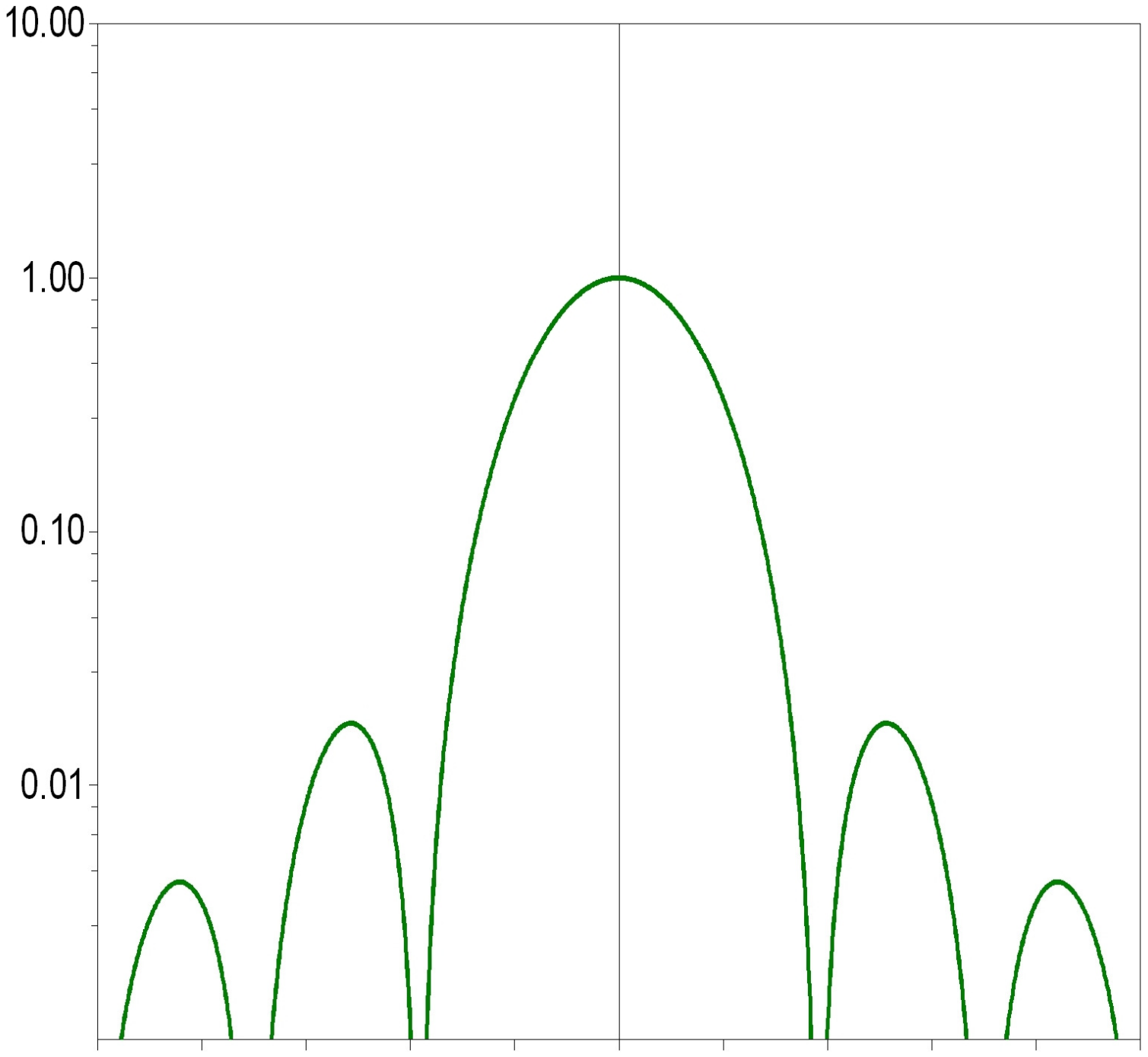

the aperture (the telescope in this case). The outer rings of the pattern are very faint, so essentially the diameter of a star image is just twice the Airy Disk radius.

In theory all stars have the same diameter image because all have the same Airy Disk radius. However, the star image diameter seen by a telescope user like Galileo depends not just on the Airy Disk radius, but also on factors that set a limit on the intensity of light that can be detected, such sky conditions and the sensitivity of the human eye.[10] This detection limit means that the apparent star diameters Galileo sees will be smaller than twice the Airy Disk radius (Figure 2). What's more, dimmer stars will appear to have smaller diameters than brighter stars (Figure 3). Lastly, the magnitude scale Galileo used (still used today) for measuring star brightness is logarithmic. A first-magnitude star is 2.512 times as bright as a second-magnitude star, which is 2.512 times as bright as a third-magnitude star, and so forth. Calculating the relationship between diameter and magnitude yields a relationship that is, in fact, fairly linear, and would doubtlessly look linear to Galileo, working at the limits of what his eyes and instrument can do (Figure 4).

In conclusion, diffraction tricked Galileo into believing that a linear relationship existed between the magnitudes and apparent sizes of stars, and therefore (since he assumed stars were suns) between the magnitudes and distances of stars. Since an understanding of wave optics lay almost two centuries in Galileo's future[11], he can certainly be forgiven for not grasping that diffraction was creating spurious results! Galileo's method for calculating the distances to stars made perfect sense, but for diffraction's trickery. Thus diffraction tricked Galileo into mismeasuring the distances to the stars, and to this physics teacher, that's a pretty interesting historical anecdote to share with students.

FIGURE 2

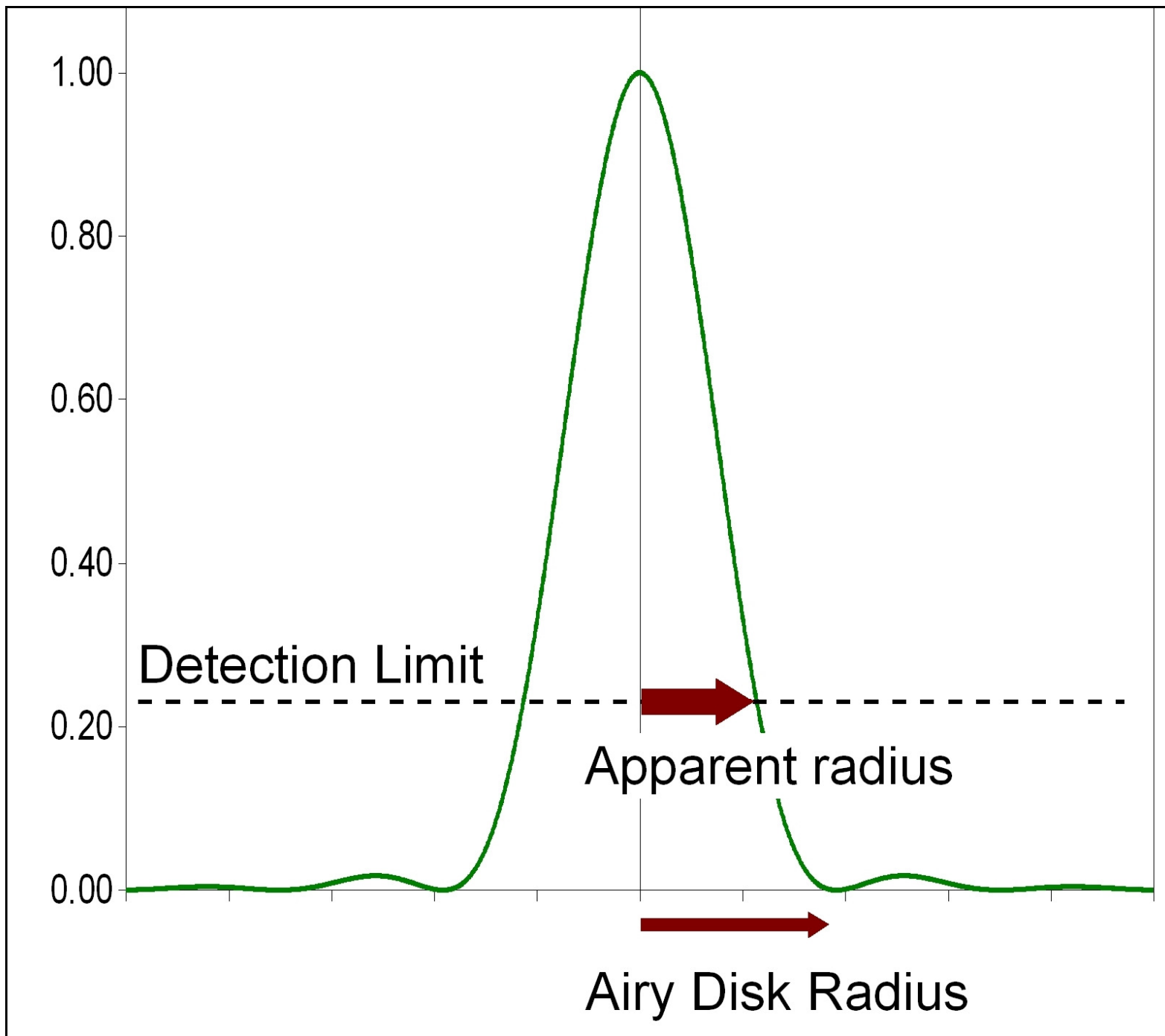

**Factors such as sky conditions and the sensitivity of the human eye mean that there exists a detection limit below which Galileo's eye cannot detect a star's light. The apparent radius of a star image is less than the Airy Disk radius due to the intensity curve of the star's diffraction pattern falling below this detection limit. Galileo can only see that part of the diffraction pattern where the intensity is greater than the limit, and thus the star appears smaller than its Airy Disk radius.**

FIGURE 3

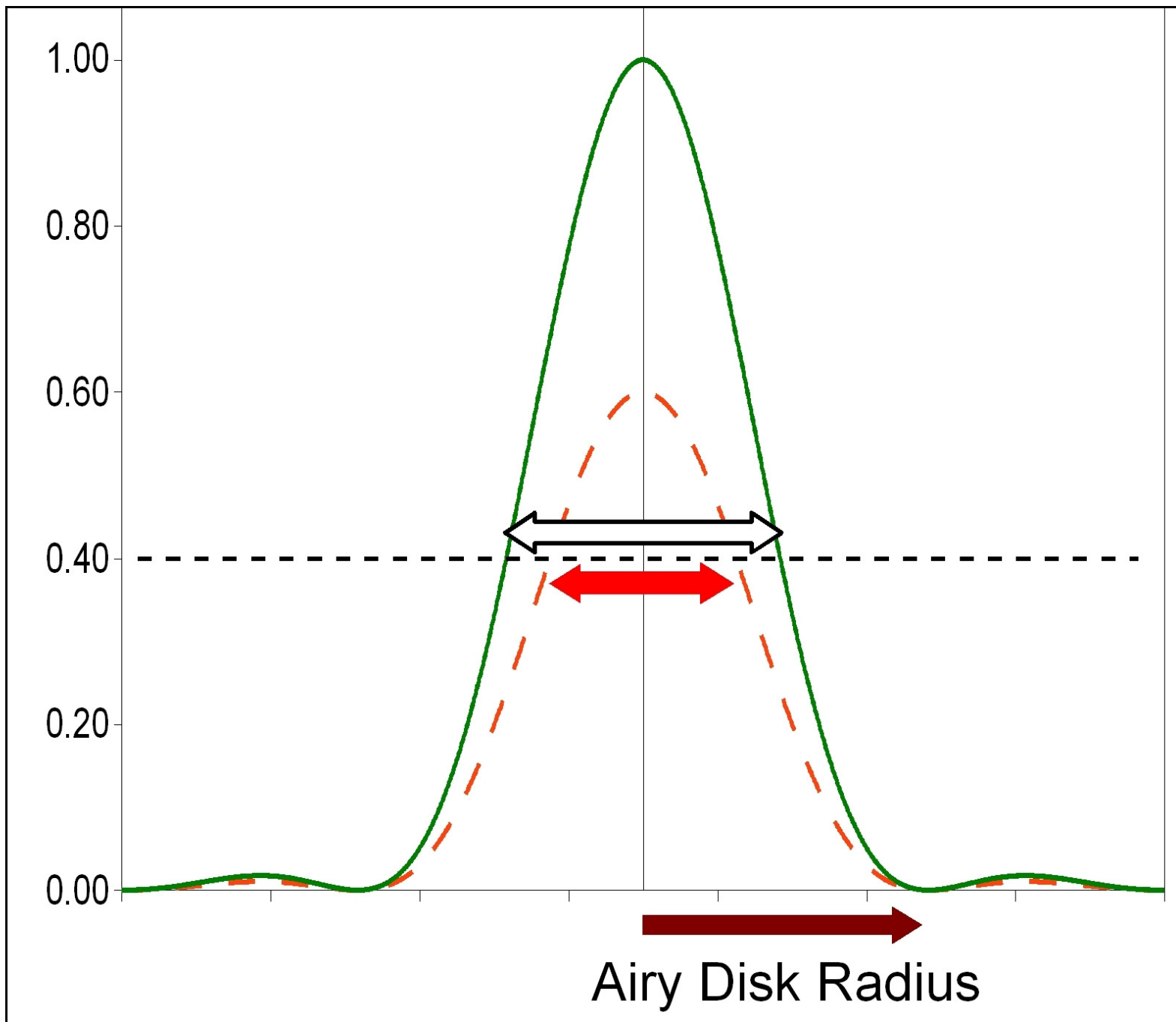

**Diffraction pattern intensity plots for two stars, the dimmer (dashed plot) with intensity 60% of the brighter (solid plot). Both stars have same Airy Disk radius, but due to the detection limit the brighter star has a larger apparent diameter (topmost arrow) than the dimmer star (next arrow down). The detection limits in this and the previous figure are chosen for clarity of illustration and are not necessarily representative of any particular combination of sky conditions and eye sensitivity.**

FIGURE 4

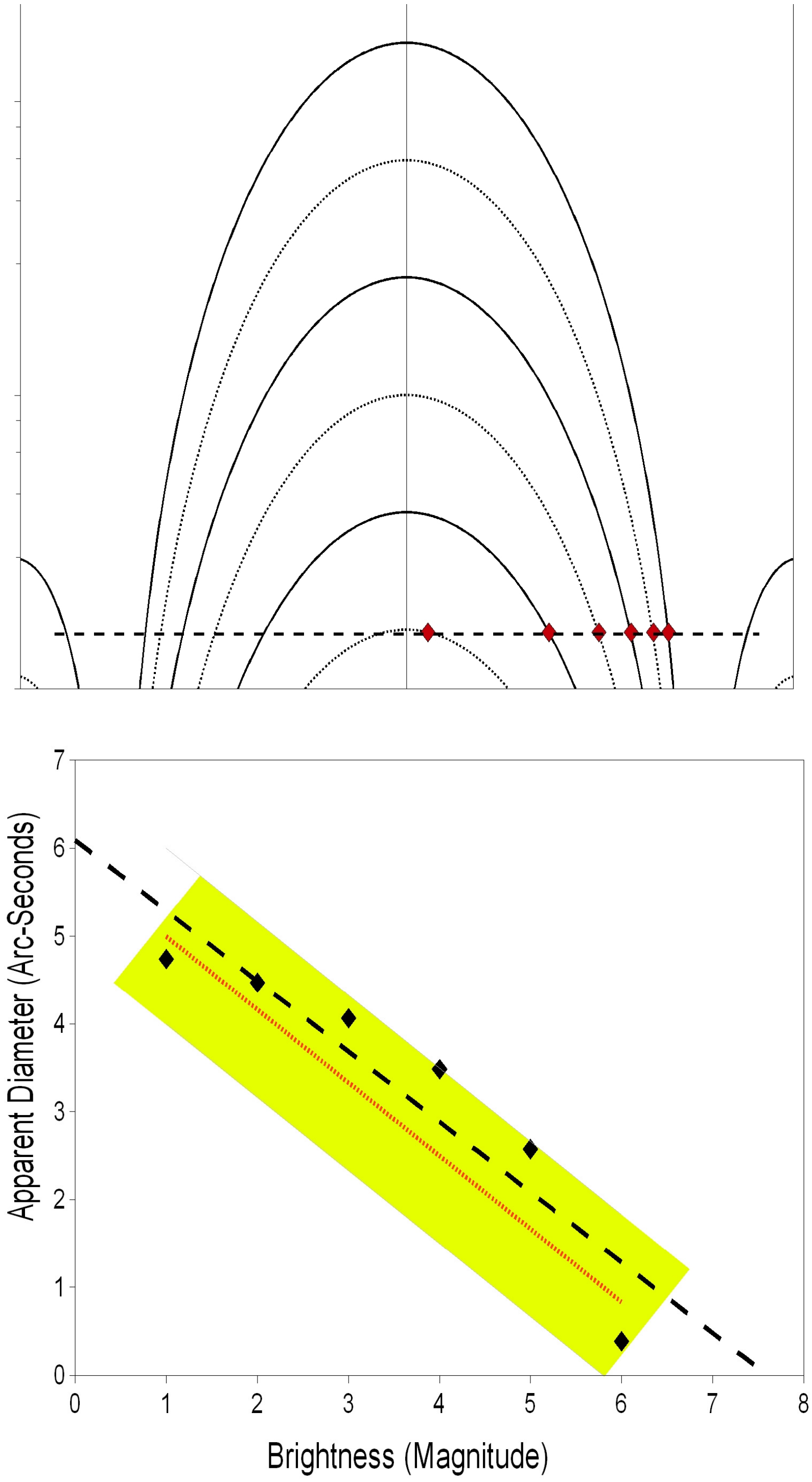

**Top -- Intensity plots (log axis) for stars of magnitudes 1-6. The size of each star's image (marked by diamonds) is determined by where the diffraction intensity curve drops below the detection limit.**

**Bottom -- Plot of apparent star diameter vs. magnitude. Finely dashed red line represents relationship between magnitude and size given by Galileo, with yellow zone showing one arc-second "error" to either side. Points are star diameters as calculated in Figure 4A. Coarsely dashed black line is fit to the points. If Galileo is limited by an error of an arc-second in his measurements then his observations will not contradict his theory of a linear diameter-vs.-magnitude relationship.[12]**